\documentclass[a4paper,11pt]{article}
\usepackage[a4paper,left=3cm,right=3cm,top=3cm,bottom=2.5cm]{geometry}
\usepackage[utf8]{inputenc}
\usepackage{amsmath,amssymb}
\usepackage{graphicx}
\usepackage{color}
\usepackage{cite}
\usepackage{listings}

\usepackage[T1]{fontenc}
\usepackage{hyperref}

\usepackage{sourcecodepro}

\definecolor{codepurple}{rgb}{0.651, 0.0149, 0.643}
\definecolor{codegray}{rgb}{0.3,0.3,0.3}
\definecolor{numbergray}{rgb}{0.5,0.5,0.5}
\definecolor{codeblue}{rgb}{0.251, 0.471, 0.949}

\begin{document}
{\flushright LMU-ASC 32/22\\
}\vspace{2cm}
\begin{center}
\includegraphics[width=0.3\textwidth]{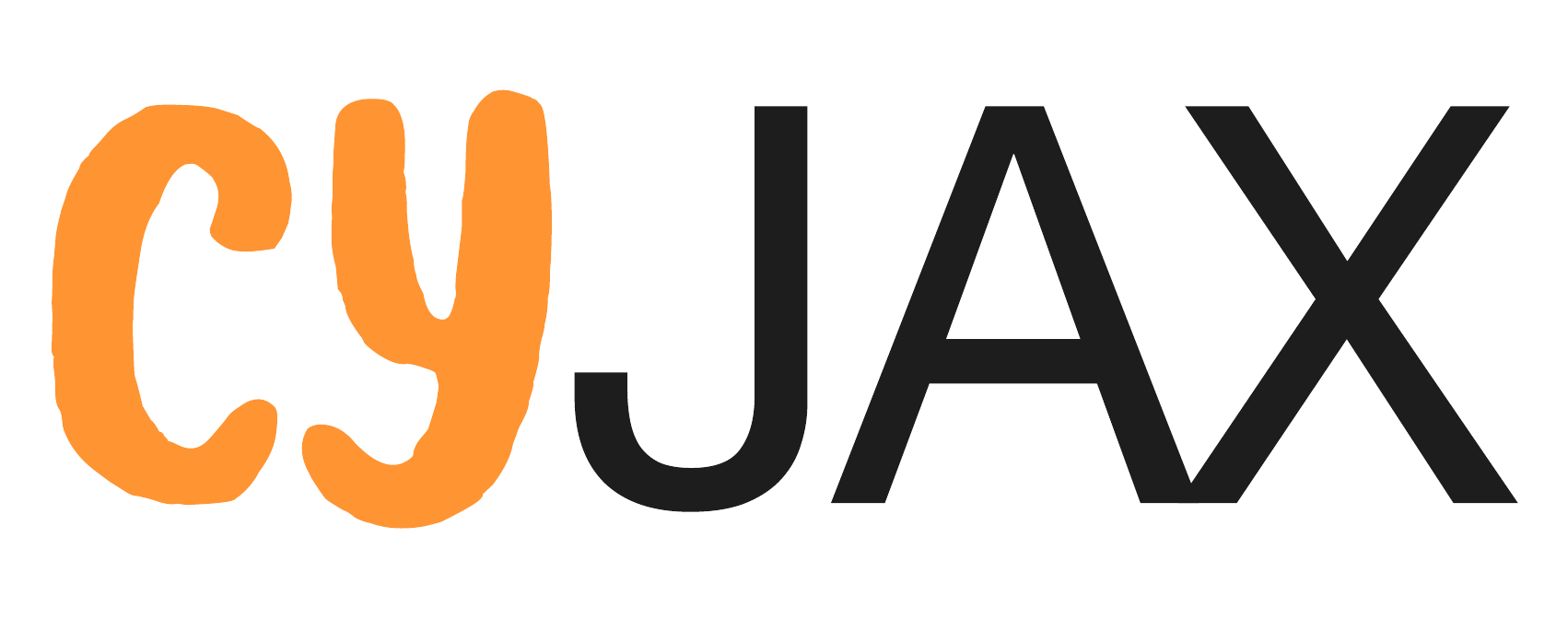}~\\[0.3cm] {\LARGE \bf A package for Calabi-Yau metrics with JAX}\\[0.5cm]
{Mathis Gerdes$^{1,*}$, Sven Krippendorf$^{2,3,*}$}\\
{\it
$^{1}$ Institute of Physics, University of Amsterdam, the Netherlands\\
$^{2}$ Universit\"ats-Sternwarte, LMU Munich, Germany\\
$^{3}$ Arnold Sommerfeld Center for Theoretical Physics, LMU Munich, Germany\\
$^{*}$ Authors to whom any correspondence should be addressed.}\\[0.5em]

{\bf E-mail:} m.gerdes@uva.nl, sven.krippendorf@physik.uni-muenchen.de
\end{center}

\begin{abstract}
\noindent We present the first version of CYJAX, a package for machine learning Calabi-Yau metrics using JAX. It is meant to be accessible both as a top-level tool and as a library of modular functions. CYJAX is currently centered around the algebraic ansatz for the Kähler potential which automatically satisfies Kählerity and compatibility on patch overlaps. As of now, this implementation is limited to varieties defined by a single defining equation on one complex projective space. We comment on some planned generalizations.\\[0.2cm] \noindent More documentation can be found at: \url{https://cyjax.readthedocs.io}.\\
\noindent The code is available at: \url{https://github.com/ml4physics/cyjax}.

\end{abstract}
\newpage
\tableofcontents

\section{Introduction}
Calabi-Yau (CY) manifolds appear ubiquitously in the context of string theory as extra-dimensional geometries, since a large fraction of consistent string theory 
constructions 
is based on such CY-manifolds.
The geometry of these manifolds determines the resulting effective field theories (EFT). Unlike topological properties, such as the particle spectrum, which are accessible with existing mathematical tools, the metric itself is generally unknown for these spaces. However, this provides a gap to connect string theory with its low-energy observables such as the couplings among these particles which are determined by the metric. 
In the absence of analytic solutions, several numerical approaches have been proposed to identify metrics 
of CY-manifolds. In short, there are currently two approaches which can be used to approximate CY metrics:
\begin{itemize}
\item Using a 
fixed-point 
iteration scheme which is widely referred to as Donaldson's algorithm~\cite{Donaldson:2005mat,Douglas:2006hz,Douglas:2006rr,Braun:2007sn,Braun:2008jp,Ashmore:2019wzb}. This spectral method was developed after previous approaches using finite difference methods appeared unpractical~\cite{Headrick:2005ch}.
\item Using energy functional minimization where the energy functional measures 
the Ricci flatness or equivalent quantities such as one given by an appropriate Monge-Amp\`ere loss~\cite{Headrick:2009jz,Anderson:2020hux,Douglas:2020hpv,Jejjala:2020wcc,Larfors:2022nep,Berglund:2022gvm}. At the moment, this approach is computationally the most efficient.
\end{itemize}
Obtaining either of the measures involves specialized code which has been developed for several proof of concept examples. To facilitate the development of these methods and their application to other manifolds, we are developing the package CYJAX which aims at providing a modular environment that addresses 
extra-dimensional 
metrics, in particular CY metrics. 
The modularity of the implementation makes it possible to extend and adapt the package for dedicated use cases. 
Our code is based on JAX~\cite{jax2018github} which enables significant acceleration from pure python code via the use of \texttt{jit}-compilation and automatic vectorization using \texttt{vmap} \footnote{\texttt{jit} stands for just-in-time compilation, which means that the python code is compiled into optimized algebraic instructions for the CPU/GPU which can dramatically outperform pure python code.
The \texttt{vmap} function allows efficient implementation of algorithms which operate over batches of data at once. For more details, see e.g.~the JAX documentation\cite{jax2018github}.}.  
It also allows programmatic generation of efficient code starting from symbolically derived expressions, e.g.~using SYMPY.
This article is a short note on the first release of CYJAX.

The basis of this code was developed in large parts for the Master thesis of one of us~\cite{mathismaster} and has been used for some experiments 
in~\cite{Anderson:2020hux}.
This package is complementary to the package CYMETRIC~\cite{Larfors:2022nep}, making different code design choices and using a different ansatz for the metric. Instead of directly parametrizing it by a neural network, we use an algebraic ansatz which is guaranteed to be Kähler and which always matches on patch overlaps. Compared to CYMETRIC, we can thus use two fewer losses.
Further, we focus immediately on learning moduli-dependent approximations of the Calabi-Yau metric, instead of optimizing for single points in moduli space. 
Beyond this, we hope that independent implementations of several algorithmic aspects may be useful to the community. 
In future work, we expect direct quantitative comparisons to be straight-forward as JAX (used by us) and Tensorflow/NumPy (used in CYMETRIC) are interoperable to a large extent. 
This may also allow using aspects from each implementation for a particular task.
Importantly, our package provides an independent 
determination of the metric which can be useful in comparing the performance.\footnote{For the high energy theorists, this has been useful in the 
past to improve the quality of predictions of supersymmetric mass spectra where different spectrum generators displayed discrepancies in the 
predicted Standard Model Higgs mass~\cite{Allanach:2018fif,Allanach:2017hcf}.}

The rest of this note is organized as follows. Section~\ref{sec:varieties} introduces the manifolds, our choices for relevant quantities to study metrics, and comments on their respective code implementation. Section~\ref{sec:tutorials} showcases a few examples on how CYJAX can be used. In Section~\ref{sec:outlook} we provide an overview of some of the planned functionalities for future releases.

\section{Choices for CY Manifolds}
\label{sec:varieties}
We consider here varieties $X \subset \mathbb{P}^{d+1}$ which are defined as the zero-locus of a single homogeneous polynomial:
\begin{equation} 
Q(z) = \sum_{\alpha} \psi_\alpha z^\alpha \Bigg|_{z \in X} = 0 \,.
\end{equation}
Recall that the {\bf homogeneous coordinates} $z=[z_0: \ldots: z_{d+1}]$ on projective space $\mathbb{P}^{d+1}$ are just the complex coordinates of $\mathbb{C}^{d+2}$ with the identification $z \sim \lambda z$ for all $\lambda \in \mathbb{C} \setminus \{0\}$.
The multi-index $\alpha$ ranges over all natural numbers such that $\sum_{i=0}^{d+1} \alpha_i = d+2$, corresponding to all monomials of degree $d+2:$
\begin{equation}
z^\alpha = \prod_{i=0}^{d+1} z_i^{\alpha_i} \,.
\end{equation}
Choosing particular values of coefficients $\psi_\alpha \in \mathbb{C}$ corresponds to choosing the complex structure moduli.

The dimension of the variety $X$ in the above example is $d$. 
Of particular interest in physics is the case $d=3$, where $X$ is called a quintic threefold (referring to the degree of the defining equation and manifold's dimension, respectively).
We may restrict which coefficients $\psi$ we allow to be nonzero.
A particular example is the so-called Dwork family of quintics, which has a single complex parameter:
\begin{equation}
0=\sum_{i=0}^{4} z_i^{5} - 5 \psi \prod_{i=0}^{4} z_i \,.
\label{eq:dwork}
\end{equation}
Note that in this case we have dropped the index on $\psi$ as there is only one nonzero component for $\alpha=(1,\ldots,1)$.

One can show that the first Chern class vanishes for the above manifolds (see e.g.~\cite{Greene:1996cy}).
From the work by Yau~\cite{Yau:1978cfy}, we thus know that a Ricci flat-metric exists and we want to find a numerical approximation to it.
The present case of examples has a single K\"ahler modulus, $h_{11}=1$, so the metric's dependence on this modulus is absorbed by an appropriate rescaling. 
A generalization to the case with multiple K\"ahler moduli is left for the future.

In CYJAX, varieties of the above type are represented by the \texttt{VarietySingle} class, which are essentially characterized by their defining polynomial. There are multiple ways to specify this polynomial in code, the most convenient of which is to pass it as a SYMPY-style string expression via \texttt{VarietySingle.from\_sympy}. For convenience, the Dwork and Fermat quintic are also readily implemented (\texttt{Fermat}, \texttt{Dwork}).

\subsection{Choice of coordinates}
To study the metric on these varieties, we have to identify suitable coordinates. Below, we describe the choices on how to relate the coordinates on the ambient space with coordinates on the variety.
There is no unique choice for going from the $d+2$ coordinates in the ambient complex space to proper coordinates on the $d$-dimensional variety~$X$. 
One option is to accept the redundancy and use the full set of {\bf homogeneous coordinates}. However, some geometric quantities have no numeric globally defined representation, and thus a choice about local coordinate patches has to be made. The description below also serves as an overview of the conventions we choose in CYJAX.

\subsubsection{Coordinates in ambient space}
To remove the scaling ambiguity in homogeneous coordinates, we pick one index (with non-zero entry) and set its value to $1$ by rescaling:
\begin{align}
\nonumber [z_0: \ldots: z_p:  \ldots: z_{d+1}]
&\sim [z_0/z_p: \ldots: 1:  \ldots: z_{d+1}/z_p] \\
&\equiv (z_0/z_p, \ldots z_{d+1}/z_p)
= (z^{(p)}_0, \ldots, z^{(p)}_{d}) = z^{(p)} \,.
\end{align}
If the \textbf{patch} index $p$ of the $d+2$ homogeneous coordinates is scaled to one and omitted, we denote the remaining $d+1$ {\bf affine coordinates} by $z^{(p)}$.
Computationally, the affine coordinates are represented by an array with $d+1$ entries together with an integer specifying the patch they are in (i.e.~which homogeneous index was scaled to $1$).

Going from homogeneous coordinates to patch $p$, we divide by the value of $z_p$.
Numerically, it is advantageous to avoid very large values.
For the numerically ``optimal'' patch we thus, by default, choose $p$ such that $|z_p|$ is maximal. We still have two choices of how to store local coordinates:
\begin{enumerate}
\item We can keep the full array of homogeneous coordinates but always scale the largest value to $1$.
   This allows each coordinate to be represented uniquely by a single array.
   However, since the patch index is implicit, we could then not force a function to treat numerical inputs as lying in another patch.
\item  Alternatively, we can keep an array of affine coordinates together with a patch index.
   This has the benefit of saving slightly in computational cost, especially where the patch index has to be
   known explicitly, at the memory cost of carrying around two arrays.
\end{enumerate}
Most functions implemented in CYJAX can be called both with homogeneous and local coordinates on projective space.
The cases are distinguished by whether or not a patch index is supplied.

When indexing into affine or homogeneous coordinate arrays one has to be somewhat careful, as removing the `redundant' element with value 1 shifts indices.
The index $k$ with respect to the affine coordinates $z_k^{(p)}$ will here be referred to as {\it affine} index.

\subsubsection{Local coordinates on the variety}

Going from affine coordinates to coordinates on the variety $X$ itself we must eliminate one additional redundant entry.
Given coordinate values $z_1^{(p)}, \ldots, z_{d}^{(p)}$ we can recover $z_0^{(p)}$ by solving the defining equation.
The simplest way to pick coordinates is by choosing one coordinate index that is to be considered ``dependent'' on the other values. 
All other coordinate entries are kept.

As a particular example, consider a $(d+1) \times (d+1)$ matrix $g$ denoting a metric in the ambient projective space.
We now want to compute the pullback of this to the variety.
We have
\begin{equation}
g^X_{i\bar{\jmath}} = \frac{\partial z_p^{k}}{\partial z_p^{i}} \frac{\partial \bar{z}_p^{\bar{l}}}{\partial \bar{z}_p^{\bar{\jmath}}}~g_{k\bar{l}} \,,
\end{equation}
where $i$ and $\bar{\jmath}$ only range over $d$ values corresponding to the choice of dependent entry. The induced metric can be calculated via \texttt{induced\_metric}.
If $z_p^{m}$, i.e.~index $m$, is chosen as dependent variable, only $\partial z_p^{m}/\partial z_p^i$ is non-trivial (all others being either one or zero).
The Jacobians can be computed directly (and automatically) from the defining equation $Q(z)=0$:
\begin{equation}
\frac{dz_{m}}{dz_j} = - \frac{dQ}{dz_j} \left( \frac{d Q}{dz_{m}} \right)^{-1} \,.
\end{equation}
This is implemented using the function \texttt{jacobian\_embed}. Inside the respective variety we work with the ``optimal'' dependent coordinate which is minimizing the entries in the Jacobian. 

\subsection{Sampling points}
\label{sec:sampling}
To numerically evaluate integrals, e.g.~to compute the volume, we use a Monte Carlo approximation.
We thus need to generate samples that lie on the manifold and have a known distribution.
One straight-forward way of generating points on the $d$-dimensional manifold is to sample $d-1$ random complex numbers and solve the defining equation for the last coordinate value.
However, we do not a priori know the distribution of these points.
Instead, we can sample points as intersections with a line in ambient projective space.
The distribution of these points is known~\cite{Douglas:2006rr}.

\subsubsection{Complex and projective coordinates}
If complex values are generated by sampling real and imaginary parts uniformly, points will lie on a square in the complex plane.
Uniformly sampling the radius in $[0, 1]$ and the complex angle in $[0, 2\pi]$ will ensure points lie on the unit disk, however the distribution does not have uniform density.
Instead, one should sample uniformly from the disk in $\mathbb{R}^2$ and interpret the coordinates as real and imaginary parts.

In order to generate uniformly distributed samples on $\mathbb{P}^n$, we can first sample uniform points on the real sphere $S^{2n+1}$ represented by $2n+2$ real numbers.
By pairing these up into $n+1$ complex values we obtain homogeneous coordinates on the projective space.
This construction corresponds to $\mathbb{P}^n \cong S^{2n+1} / U(1)$.
Generating uniform points on the real sphere $S^{2n+1}$ can be done efficiently by independently sampling $2n+2$ real numbers from the unit-covariance normal distribution and dividing by their vector norm.
Since the normal distribution factorizes as $p(z) \propto \exp(- \sum_i z_i^2)=\prod_i \exp(-z_i^2)$, we can efficiently draw from the joint distribution by sampling each component $z_i$ independently. The exponent $\sum_i z_i^2$ is manifestly invariant under $SO(2n+2)$ rotations, as desired for a uniform distribution, but the norm $|z|$ is not yet fixed to 1. Dividing the sampled points by their norm puts them on the unit sphere while preserving the $SO(2n+2)$ symmetry, which therefore gives us a uniform distribution on $S^{2n+1}$.

\subsubsection{Points on varieties}

To numerically estimate integrals, we make use of the Monte Carlo approximation
\begin{equation}
\int_X f d\mathrm{vol} = \int_X f \frac{d\mathrm{vol}}{dA} dA \approx \frac{1}{N} \sum_{a=1}^{N} f(x_a) w(x_a) \,.
\end{equation}
Here, $x_a$ are drawn using some (pseudo-) probabilistic procedure with known density measure $dA$ which is explicitly determined using the determinant of the induced Fubini-Study metric on the variety, i.e.~$dA={\rm det}(i^\star_p g_{\rm FS})$. 
The volume $d\mathrm{vol}$ is evaluated via the holomorphic top form which is known a priori \cite{Candelas:1987kf}.
The weights $w=d\mathrm{vol}/dA$, in the final step, are required to correct for the difference in the measures.

The sampling method we use here is described in~\cite{Douglas:2006rr}.
After uniformly sampling two points $p, q \in \mathbb{P}^{d+1}$, we can define a line $p + t q$ with $t\in\mathbb{C}$.
Samples on the variety are then given by the intersection of this line with the variety, i.e. by solutions for $t$ such that $Q(p+tq)=0$.
The density of samples generated in this way is known and given in terms of the Fubini-Study metric.

In our implementation, the variety object contains methods which allow the calculation of samples using this intersection method (\texttt{sample\_intersect}) and their weights (\texttt{sample\_intersect\_weights}).
The simplest associated application is to compute the volume via the Monte Carlo approximation (\texttt{compute\_vol}).

\subsection{Algebraic metrics}
The primary quantity we try to learn is the Hermitian matrix $H$ in the following algebraic ansatz for the Kähler potential, taken from Donaldson's algorithm:
\begin{equation}
K(z, \bar{z}) = \frac{1}{\pi k} \log \left( \sum_{\alpha\bar{\beta}} s_\alpha(z) H^{\alpha \bar{\beta}} s_{\bar{\beta}}(\bar{z}) \right) \,.
\label{eq:ansatz}
\end{equation}
Here, the $s_\alpha$ represent a set of homogeneous polynomials of some chosen degree $k$, which correspond to sections of a line bundle. Geometrically, one can understand a set of $N_k$ sections $s_\alpha$ as defining an embedding of the variety into the complex projective space of dimension $N_k-1$. 
The resulting Kähler metric on the variety can then be understood as as the pullback of the higher-dimensional Fubini-Study metric generalized by the Hermitian matrix $H$ \cite{Donaldson:2005mat}.
The larger the polynomial degree $k$, the larger the complete set of sections $N_k$ and thus the higher the potential resolution of the algebraic ansatz becomes. For Donaldson's algorithm, there is an additional requirement that the set has to form a basis of line bundle sections on the variety, because the algorithm involves a matrix inversion which is otherwise ill-defined \cite{Donaldson:2005mat}.
Any linear combination proportional to the defining polynomial vanishes and thus the full set of homogeneous monomials, for example, does not always form a basis on the variety.
For the machine learning application, however, there is no such requirement and we can use the full set of monomials of some chosen degree in ambient projective space, which merely amounts to an over-parameterization.
Note, also, that one could use any other set of linear combinations of monomials for the sections.
While both of these things can be absorbed into reparametrizations of $H$, in principle they may influence the particular training behavior and numerics.

The implementation in CYJAX explicitly exposes the choice of sections, and custom versions can be added. For convenience, two sets are implemented and can be used directly: \texttt{MonomialBasisFull} and \texttt{MonomialBasisReduced}. Finally, one needs to choose suitable initial values for the Hermitian matrix $H$ for which we present examples in Section~\ref{sec:tutorials}.

A generalization to other metric ans\"atze is left for future releases.

\begin{figure}
    \begin{center}
    \includegraphics[width=1\textwidth]{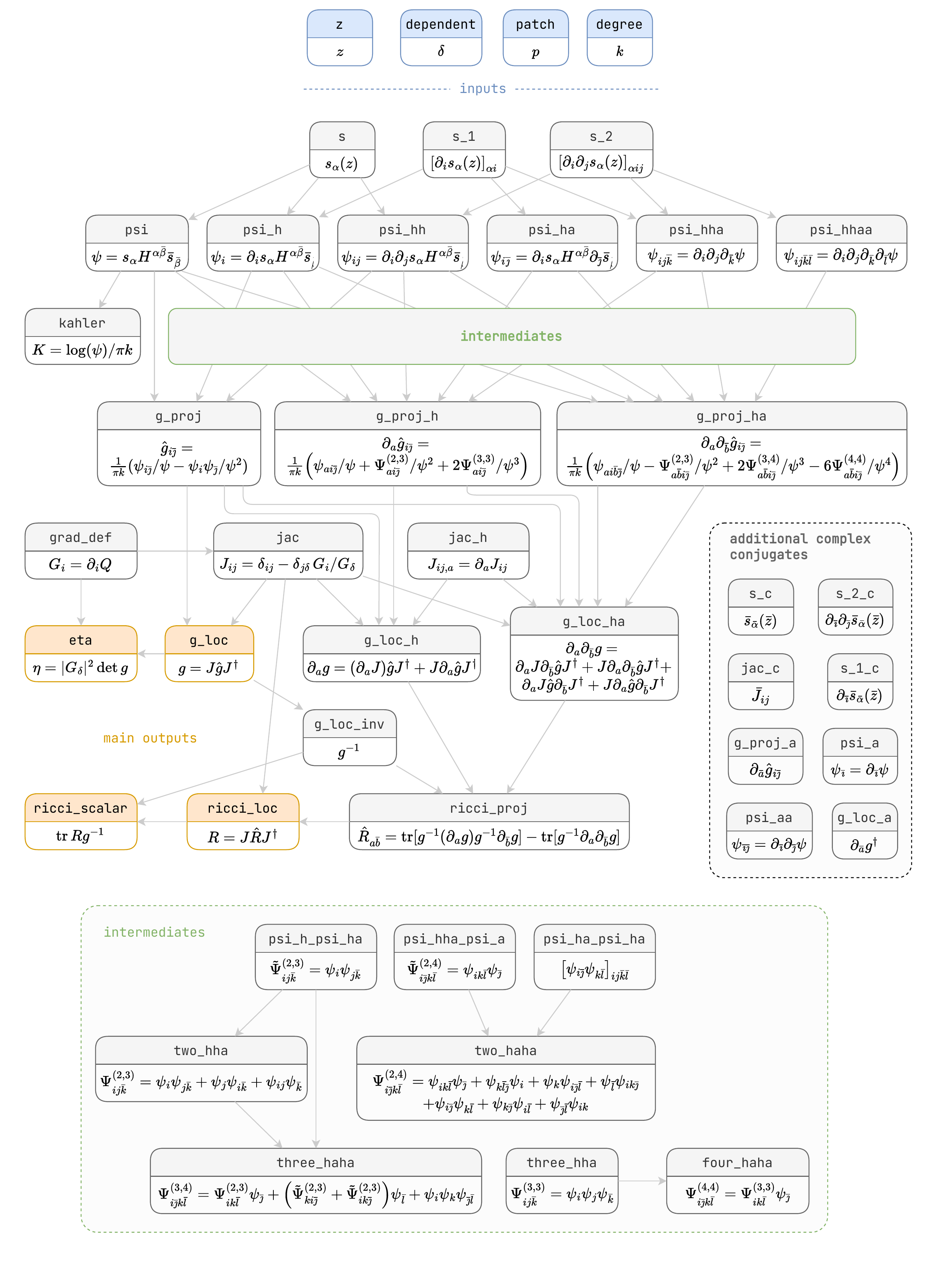}
    \end{center}
    \caption{Overview of how various geometric quantities are calculated in CYJAX. The objects correspond to the metric ansatz of equation \eqref{eq:ansatz} parametrized by the matrix $H$, computing the metric $g$ or the corresponding Kähler form $J$ for complex modulus value $\psi$ at a point on the variety specified by the homogeneous coordinate $z$. The degree of the algebraic ansatz is denoted as $z$, and the local coordinate patch on the variety is specified by the index of the affine patch $p$ and the index $\delta$ of the coordinate made dependent by solving the defining equation.}\label{fig:geometricquantities}
\end{figure}

\subsection{Geometric quantities}
Internally, the computation of geometric objects is contained in a computational graph.
This avoids duplicate code and allows easier testing of intermediate values.

Given the choice of variety, the choice of K\"ahler potential, and the ability to sample points, one is now ready to calculate various properties. Readily implemented are the calculation of: the ratio of both top-forms \texttt{eta}, the associated $\sigma-$accuracy measure \texttt{eta\_accuracy}, the Ricci curvature in local coordinates \texttt{ricci}, and the Ricci scalar \texttt{ricci\_scalar}.

Internally, some of these quantities contain intermediate steps which can also be accessed. A more detailed overview can be found in Figure~\ref{fig:geometricquantities} and in the documentation.
The computation of several quantities is implemented explicitly, but can also be computed using (repeated) automatic differentiation.
This provides a correctness check for our implementations. 
When comparing the timing (cf.~Documentation for more details), we find a slight improvement in the explicit implementation.

\subsection{Machine Learning}
The tools discussed in the previous sections can be used to implement multiple machine learning approaches to approximating the Calabi-Yau metric in any framework that works with JAX.
Functions for a particular approach to this using Flax~\cite{flax2020github} have been implemented and are accessible in the \texttt{cyjax.ml} submodule. This includes functions for initializing and working with the Cholesky decomposition of the Hermitian matrix $H$ and a batched sampler class (\texttt{cholesky\_decode}, \texttt{cholesky\_from\_param}, \texttt{hermitian\_param\_init}, \texttt{BatchSampler}).

Regarding sampling, one present constraint is that non-Hermitian eigenvalue finding is not implemented yet on the GPU. That means samples have to be generated using the CPU. We can thus either:
\begin{enumerate}
\item Train and sample on the CPU.
\item Generate and save samples to disk, then train on GPU.
\item Sample on CPU and train on GPU.
\end{enumerate}
All options are available with CYJAX.
The batched sampler mentioned above by default generates samples on the CPU and afterwards transfers them to GPU, if available.
This allows for efficient on the fly sample generation without repetition. 
The latter means the problem of overfitting, generally present in supervised learning, can be avoided.

There are multiple different losses which effectively measure the Ricci-flatness of the approximated metric.
In particular, we use here the so called $\sigma$-accuracy and a Monge-Amp\`ere loss $\mathcal{L}_{MA}$, which rely on the property that the Ricci flat metric $g$ gives rise to a volume form which must be proportional to the one given by the holomorphic top form~\cite{Headrick:2009jz}. If $\Omega$ is the holomorphic top form, we can define the ratio $\eta = \frac{\det g}{\Omega \wedge \Omega}$.
The $\sigma$ accuracy measures the deviation of $\eta$ from being constant as the integral
\begin{equation}
\sigma = \int_X |\eta - 1| d\mathrm{vol}_{\Omega} \,.
\end{equation}
For training, we use the related variance-like Monge-Amp\`ere loss
\begin{equation}
\mathcal{L}_{MA} = \sum_{z \sim X} (\eta - 1)^2 w(z)\,,
\label{eq:ma-loss}
\end{equation}
which approximates an integral with respect to the volume form $d\mathrm{vol}_{\Omega}$ using Monte Carlo weights $w(z)$.
These weights ``undo'' the bias introduced by the sampling scheme used to sample points $z$ on the manifold, as discussed in Section~\ref{sec:sampling}.

\begin{figure}[t]
\begin{center}
    \includegraphics[width=0.95\textwidth]{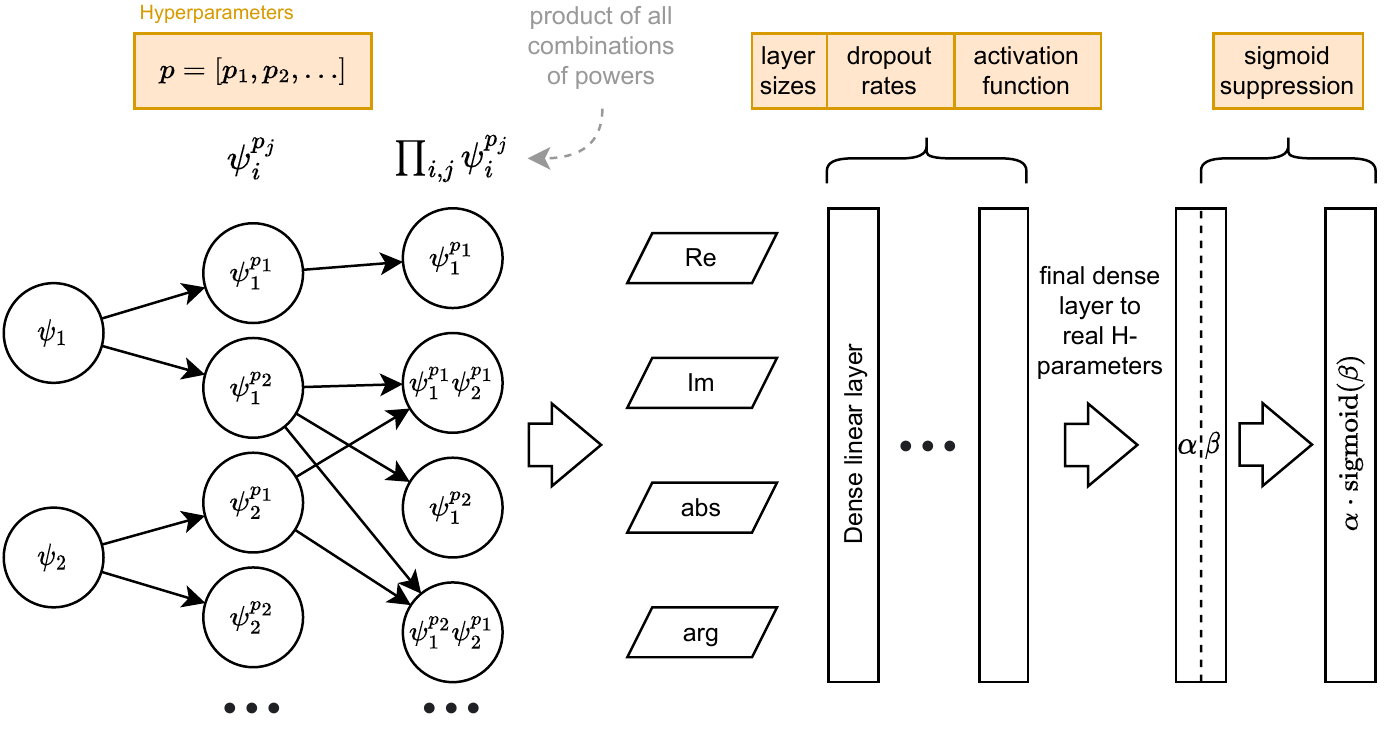}
\end{center}
\caption{Overview of MLP-Hnets and the hyperparameters, in particular the input features and the design choices for the network, which can be chosen. The network outputs parametrize the $H$ matrices in the ansatz of equation \eqref{eq:ansatz} given the moduli $\psi$ as input. The hyperparameters include fixed powers $p$ that the moduli are raised to in order to construct features which are passed to dense linear layers.  }\label{fig:Hnetwork}
\end{figure}

Lastly, there is a configurable MLP-like network for learning the moduli dependence of the $H$ matrix \texttt{HNetMLP}. For this network one can configure the features, e.g.~powers of the moduli, which can be used as input into the fully connected neural network. Several standard hyperparameters, e.g.~the layer size, activation function, can be chosen. It is beneficial to suppress almost vanishing components by multiplying them with a learnable sigmoid factor, i.e.~$\alpha\cdot{\rm sigmoid}(\beta).$ A schematic overview of the $H$ network can be found in Figure~\ref{fig:Hnetwork}. 
These networks can be easily configured as part of the configuration files for the included machine learning script.

In summary, training of the network parameters is done using the following schematic procedure which averages the loss over several moduli values in each step:
\begin{enumerate}
    \item Select $N_\psi$ moduli values $\left\{\psi^{(i)}\right\}_{i=1}^{N_\psi}$ in the desired moduli range.
    \item Apply the neural network with the current parameters $\theta$ to predict the matrices $H_\theta(\psi)$ for the selected moduli.
    \item For each $\psi^{(i)}$, sample $N_z$ points on the variety $\left\{z^{(i,j)}\right\}_{j=1}^{N_z}$ as outlined in section \ref{sec:sampling}, together with their Monte Carlo weights $w(z^{(i,j)})$.
    \item Following the computation outlined in figure \ref{fig:geometricquantities}, compute the ratio $\eta$ for each of the points and their respective moduli. Note that this ratio depends on the current approximation to the metric parametrized by $H$, and thus introduces the dependence on the model parameters $\theta$. 
    \item Evaluate the loss of equation \eqref{eq:ma-loss}, averaged over the $N_\psi$ moduli values:
    \begin{equation}
        \mathcal{L} = \frac{1}{N_\psi} \sum_{i=1}^{N_\psi} \frac{1}{N_z} \sum_{i=j}^{N_z} \left[\eta\left(z^{(i,j)}, H_\theta(\psi^{(i)}), \psi^{(i)}\right) - 1\right]^2 w\left(z^{(i,j)}\right)  \,.
    \end{equation}
    \item Compute the gradients of $\mathcal{L}$ with respect to the network parameters $\theta$ via automatic differentiation. Use these to update the network parameters by gradient descent.
    \item Repeat by returning to the first step, until some stopping criterion is met (e.g.~a fixed time elapsed, target loss reached, etc.).
\end{enumerate}
Note that $H$ parametrizes the metric over the whole variety, and thus the network itself does not take points where the metric is evaluated as input.
A detailed application of these steps to the Dwork quintic can be found in section \ref{sec:moduli-dependent-learning}.

\section{Applications}
\label{sec:tutorials}
Finally, we showcase a few applications of CYJAX. The script associated to these experiments can be found in our code repository, as well as notebooks explaining these and the procedures mentioned above.

\subsection{Moduli dependent machine learning}
\label{sec:moduli-dependent-learning}
We discuss two examples of moduli dependent machine learning. 
Readers primarily interested in the results may skip over some of the more detailed code examples.

\subsubsection{Dwork quintic with different basis resolutions}

To illustrate the explicit use of our package, we showcase here the steps to set up a neural network which learns to approximate the metric for the Dwork quintic.
Note that throughout, we have to explicitly manage a random key.
Given this state variable, pseud-random number generation is deterministic. 
In order to force careful consideration of this, and to facilitate reliable randomness in parallel computations, JAX requires explicit management of the random key.
At the beginning of a program, it is initialized by choosing some random seed, e.g.~depending on the system time or set by hand
Then, the key can be split indefinitely, and once a key has been used to generate some random values it must be discarded.

Firstly, we set up the problem by choosing the parametrized family of varieties and a monomial basis with respect to which we try to learn $H$.
\begin{lstlisting}[language=Python, label={alg:setup}, abovecaptionskip=5pt, mathescape=true]
dwork = cyjax.Dwork(3)  # three dimensional variety
degree = 5              # degree of $s_\alpha$

# Choose the whole set of monomials for $s_\alpha$.
sections = cyjax.donaldson.MonomialBasisFull(dwork.dim_projective, degree)

# Note that this is a skeleton, gathering parts that define the algebraic 
# metric and exposing various functionality; 
# it does not contain the H-matrix itself! 
# This is just like ML models in jax (e.g. flax).
metric = cyjax.donaldson.AlgebraicMetric(dwork, sections)
\end{lstlisting}
\vspace{-1cm}
The aim of our network is to learn a map $\psi \rightarrow H$ such that the corresponding algebraic metric is close to Ricci flat. For illustration, we show a simple network which only depends on the absolute value of $\psi$.

\begin{lstlisting}[language=Python, label={alg:network}, abovecaptionskip=5pt, mathescape=true]
# Next, we define the neural network as a flax module. We will optimize its 
# parameters to approximate the dependence of $H$ on the complex moduli.
class HNet(nn.Module):
    # total number of sections $s_\alpha$
    basis_size: int
    # hidden layer sizes
    layer_sizes: tuple[int] = (400, 400)
    # noise level of initialization
    init_fluctuation: float = 1e-3
        
    @nn.compact
    def __call__(self, psis):
        # psis may be 1d or 2d (if it has a batch index)
        psis = jnp.atleast_1d(psis)

        # for illustration, we only use the magnitude as input feature
        x = jnp.abs(psis).reshape(-1, 1)

        # for each hidden layer size, apply an appropriate dense layer
        for features in self.layer_sizes:
            # the first argument specifies the output size
            x = nn.Dense(features, dtype=x.dtype)(x)
            # apply a non-linear activation function
            x = nn.sigmoid(x)

        # final dense linear layer to H-parameters
        h_params = nn.Dense(
            # total number of real parameters in H
            self.basis_size**2,
            # initialize such that H starts close to the identity
            bias_init=lambda k, s, d: cyjax.ml.hermitian_param_init(
                k, self.basis_size, self.init_fluctuation),
            kernel_init=nn.initializers.constant(0.),
        )(x)

        if psis.shape == (1,):
            # remove batch dimension if input isn't batched
            return jnp.squeeze(h_params, 0)
        return h_params
\end{lstlisting}
\vspace{-1cm}

\noindent
With the neural network defined, we can now instantiate it according to the setup we have chosen above.
This is done below, followed by an example of how the model parameters are initialized and how they are used to evaluate the $H$ matrix.

\begin{lstlisting}[language=Python, label={alg:intantiate}, abovecaptionskip=5pt, mathescape=true]
# instantiate model with chosen configuration
model = HNet(metric.sections.size) 

# example value of the complex moduli (here one, so array of length 1)
psi = jnp.array([0.])
# initialize all parameters of the neural network
params = model.init(next(rns), psi)

# use parameters to evaluate neural network
h_params = model.apply(h_params, psi)
# get H-matrix from Cholesky parametrization
h = cyjax.ml.cholesky_from_param(h_params)
\end{lstlisting}
\vspace{-1cm}

\noindent
Let us now turn to the loss which is included to highlight the flexibility to change the optimisation objective. 
Here, for convenience, sampling is integrated into the loss function. 
First, we define a loss for a given fixed moduli value. 
Then we evaluate this loss over a batch of multiple moduli values which gives the loss we use per training step.

\begin{lstlisting}[language=Python, label={alg:loss}, abovecaptionskip=5pt, mathescape=true]
def eta_loss(key, psi, h_param, sample_size):
    """Compute variance-based eta loss."""
    # create samples for MC integral
    (zs, patch), weights = dwork.sample_intersect(
        key, psi, sample_size, weights=True, affine=True)
    
    # construct H matrix from Cholesky parametrization
    h = cyjax.ml.cholesky_from_param(h_param)

    # compute the eta ratios using the algebraic ansatz
    etas = metric.eta(h, zs, psi, patch).real

    # Since we only want eta to be constant, we compute
    # it's current mean and use this to normalize eta below.
    # We can stop gradient computation so this normalization
    # is ignored in the computation of gradients.
    etas_sg = jax.lax.stop_gradient(etas)
    eta_mean = jnp.mean(weights * etas_sg) / jnp.mean(weights)

    # compute and return variance-like eta loss
    loss = (etas / eta_mean - 1) ** 2
    return jnp.mean(weights * loss)
    
# for loss function, compute above for multiple random values of psi
def loss_function(params, key, sample_size=500, psi_rad=10, batches=4):
    # need to split random key since one is "consumed" to generate psis
    key, k1 = jax.random.split(key)
    # sample psis with uniform complex angle and given radius
    # shape: (batches, 1) -> (batch size, number of moduli)
    psis = cyjax.random.uniform_angle(k1, (batches, 1), 0, psi_rad)

    # apply neural network for moduli values
    h_params = model.apply(params, psis)

    # generate different random key for each value of psi
    keys = jax.random.split(key, batches)

    # vectorize the eta_loss over all but the last argument, which is fixed
    vect_loss = jax.vmap(eta_loss, in_axes=(0, 0, 0, None))
    # apply to batch and return the mean loss
    loss = vect_loss(keys, psis, h_params, sample_size)
    return jnp.mean(loss)
\end{lstlisting}
\vspace{-1cm}

\noindent
The output of the loss function is effectively an average measure (randomly chosen) over the complex moduli parameter range.
The training of our network can proceed by a standard training loop.

\begin{figure}
\begin{center}
\includegraphics[width=0.79\textwidth]{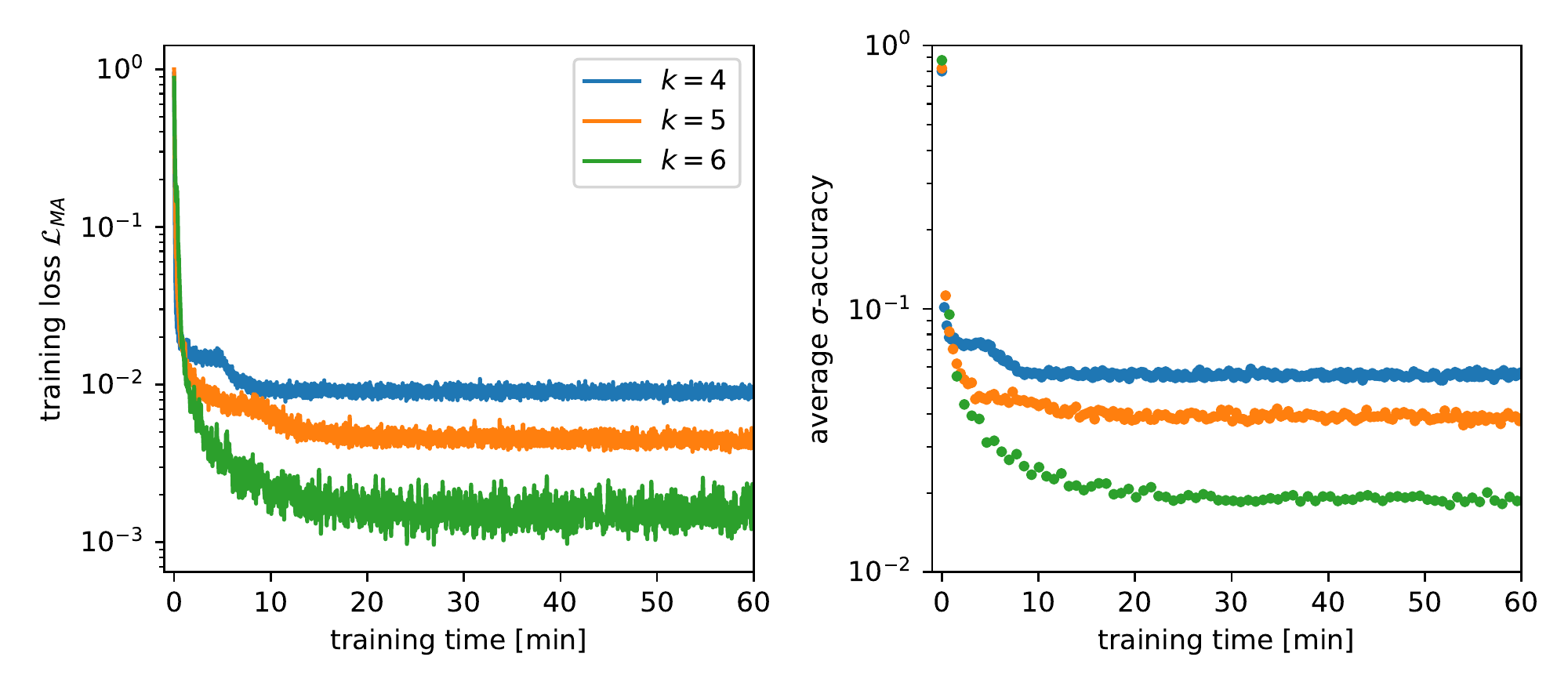}\vspace*{1em}
\includegraphics[width=0.49\textwidth]{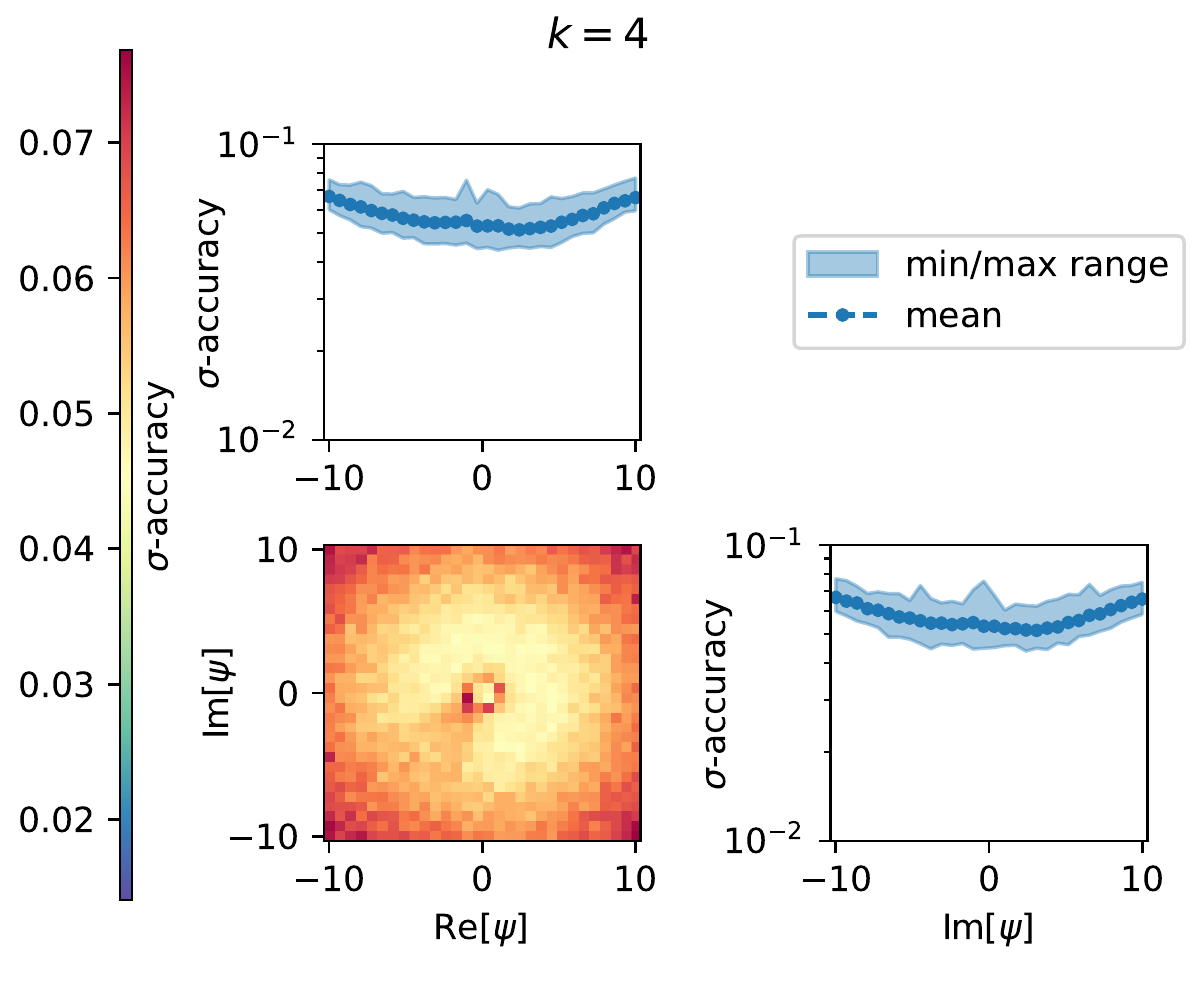}\hspace*{.015\textwidth
}
\includegraphics[width=0.49\textwidth]{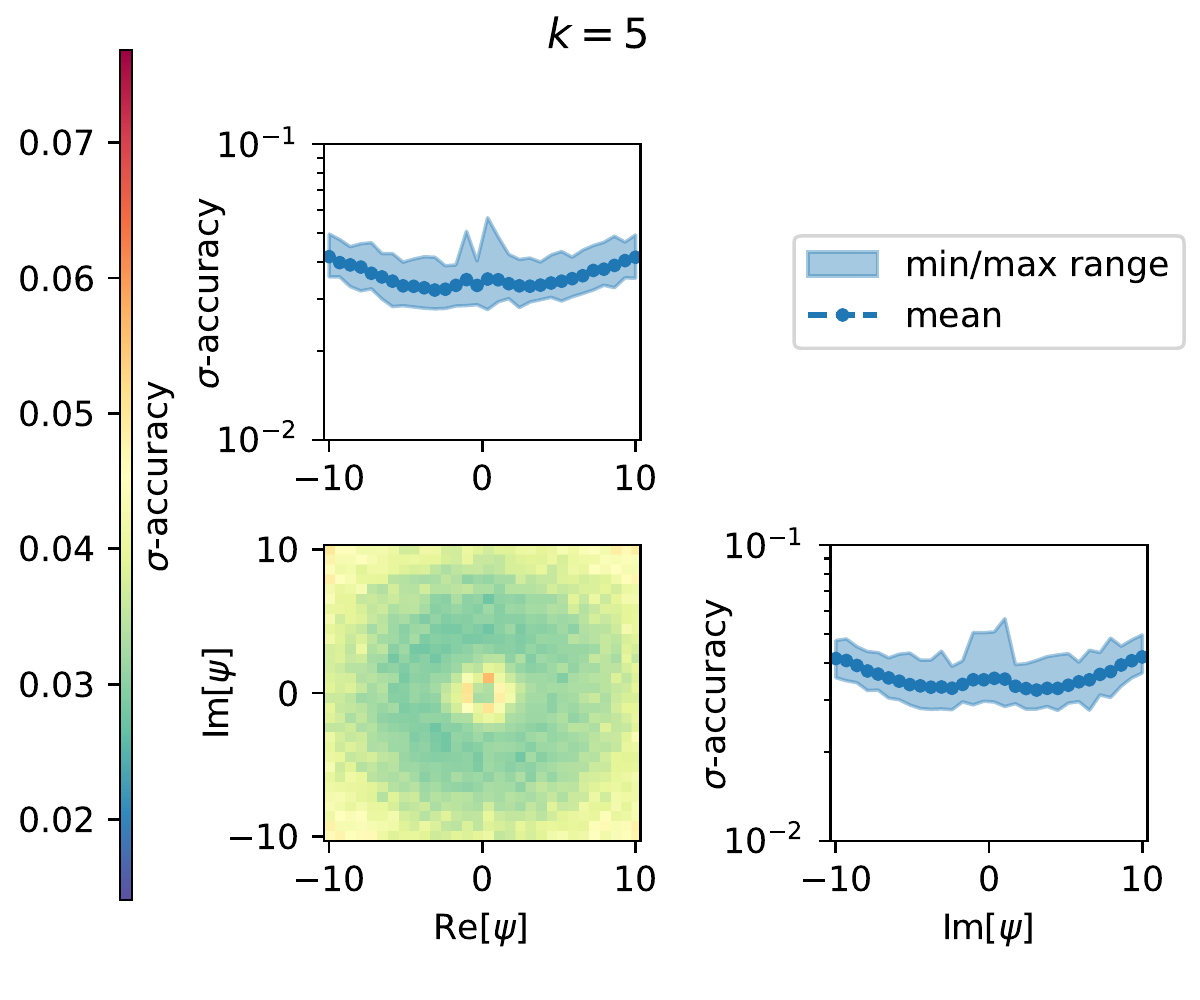}\vspace*{1em}
\includegraphics[width=0.49\textwidth]{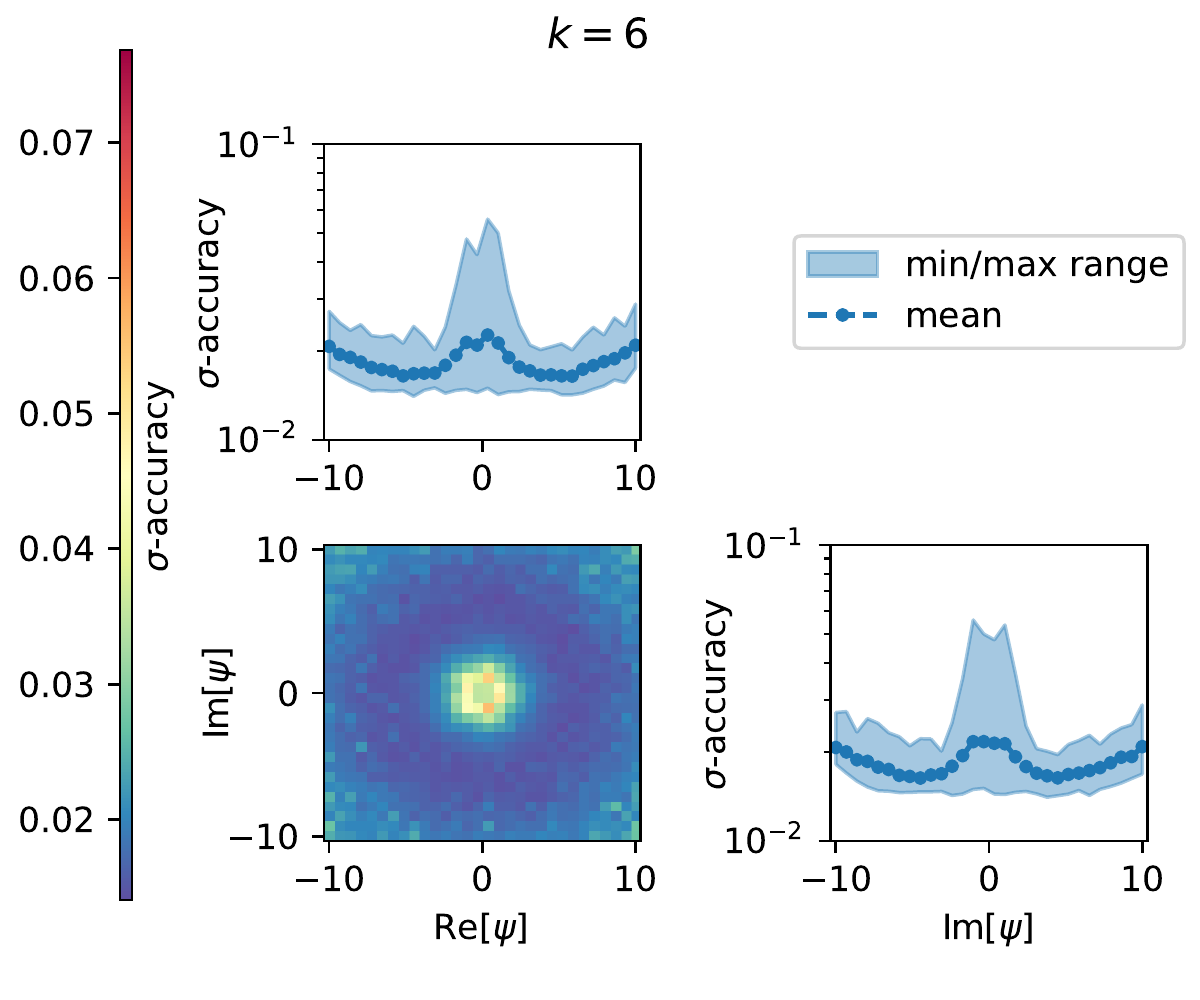}
\end{center}
\caption{The metric for the Dwork quintic with different spectral resolutions. The top row shows our training behavior and the bottom rows display the accuracy after training. The shown loss during training is a running average over $100$ training steps. Each training gradient descent step consists of 4 uniformly drawn samples of $\psi$ and, for each, $500$ points on the variety. The Adam optimizer with exponentially decaying learning rate starting at $10^{-3}$ was used. The $\sigma$-accuracy evaluated during training is the average over $10$ initially drawn and fixed moduli values. All $\sigma$-accuracies were evaluated using $1000$ independently drawn points on the variety. The edges of the final accuracy plots show values marginalized to only one real component by taking the mean, the maximum, and the minimum over the respectively other real component.}\label{fig:training-dwork}
\end{figure}

\begin{lstlisting}[language=Python, label={alg:train}, abovecaptionskip=5pt, mathescape=true]
# choose an optimizer and learning rate
opt = optax.adam(1e-3)
# initialize optimizer given network parameters
opt_state = opt.init(params)

# Below we define the update setp, which optimizes the parameters
# by stochastic gradient descent. We jit-compile this function to
# speed it up, since it will be called in each optimization step.

@jax.jit
def update_step(key, params, opt_state):
    # compute gradients for given parameters & random key
    grads = jax.grad(loss_function)(params, key)
    # determine "updates" parameters are changed by, given gradients
    updates, opt_state = opt.update(grads, opt_state)
    # change parameters by the updates
    params = optax.apply_updates(params, updates)
    # return new parameters and optimizer state
    return params, opt_state

# number of steps to train for
n_steps = 200
# pick an initial random seed
key = jax.random.PRNGKey(42)
for i in range(n_steps):
    # split random key and iterate update step
    key, key_iter = jax.ranodm.split(key)
    params, opt_state = update_step(key_iter, params, opt_state)
\end{lstlisting}
\vspace{-1cm}
Figure~\ref{fig:training-dwork} shows the results of training\footnote{All training was done on a single Titan RTX GPU, for which the training time is shown here.} for the Dwork family of equation \eqref{eq:dwork} using a network architecture as in Figure \ref{fig:Hnetwork} with two hidden layers of size $4096$, feature-powers $p=[1,2,3,4,5]$, and for degrees $k=4,5,6$.
Training was done for values in the square given by $|{\rm Re}[\psi]|, |{\rm Im}[\psi]| < 10$.
As expected, we find improved accuracies for higher resolutions.

\subsubsection{A two-moduli example}
\label{sec:twomoduli}
\begin{figure}
\begin{center}
\includegraphics[width=0.84\textwidth]{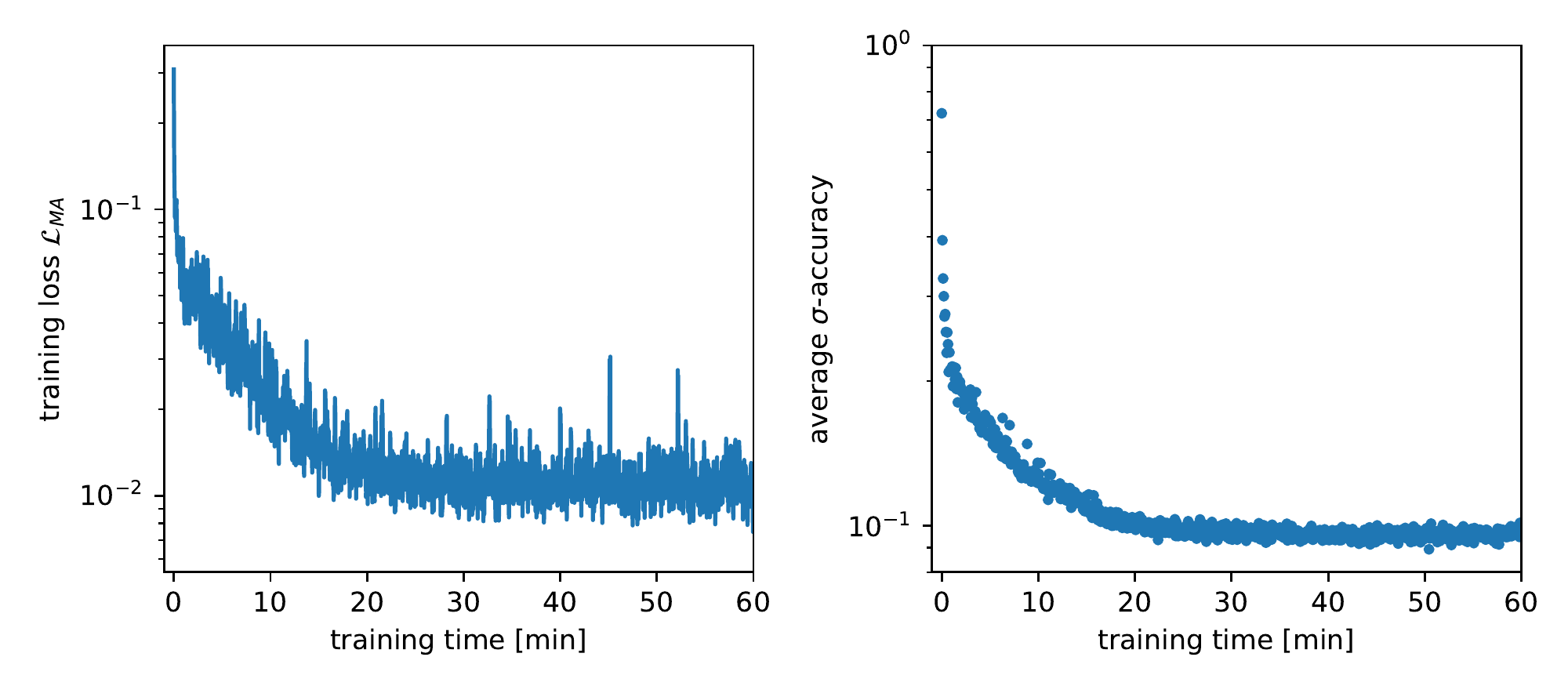}\vspace*{2em}
\includegraphics[width=0.88\textwidth]{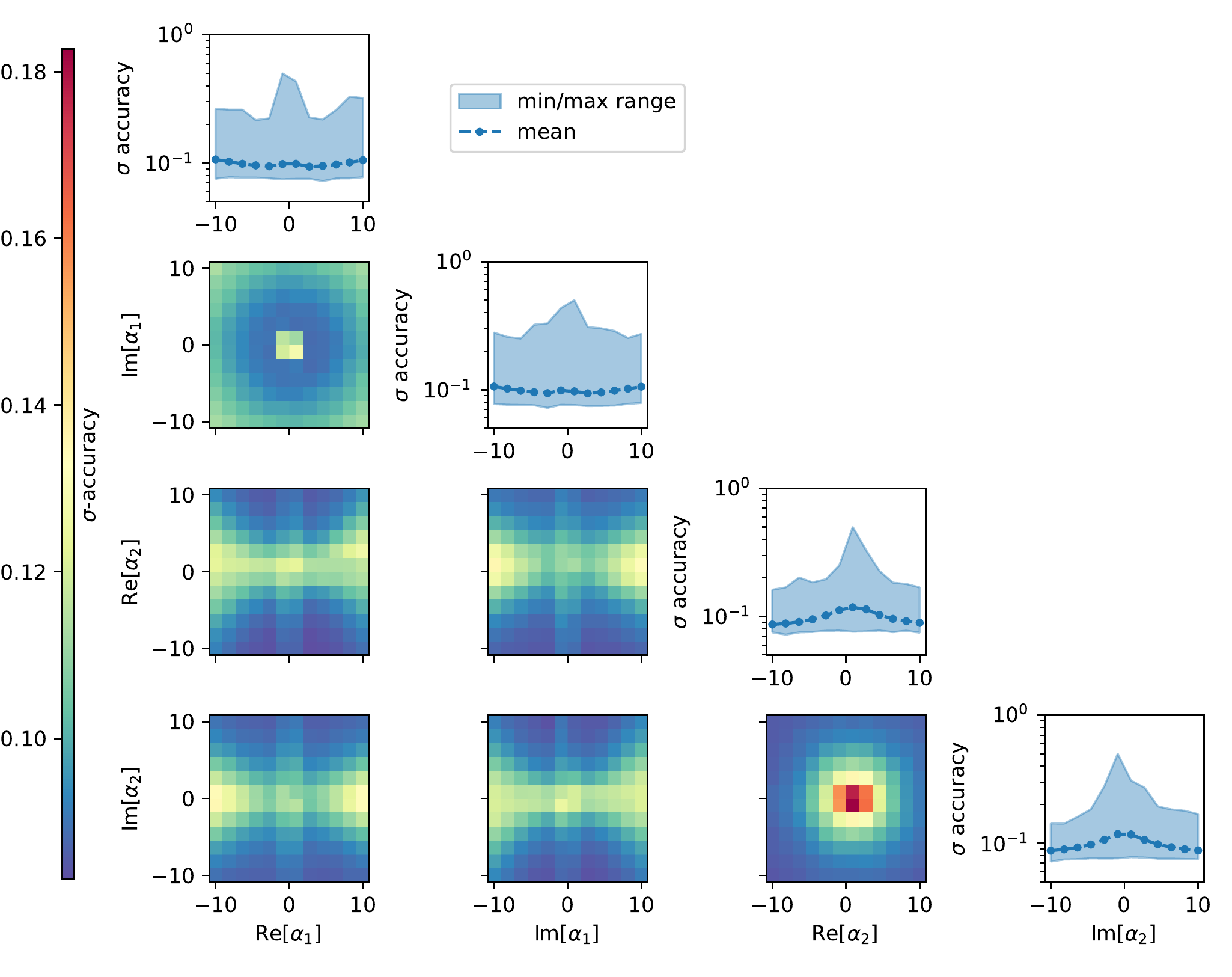}
\end{center}
\caption{
    Training behavior and obtained accuracy for our two moduli example of the quintic as defined in Equation~\eqref{eq:2moduli}. 
    Training hyperparameters are chosen as in Figure~\ref{fig:training-dwork}, except with two input moduli and powers from $1$ to $6$ as well as $10$ moduli values per training batch. 
    The marginalization for the heatmaps was done by taking the average over respectively all other real moduli parameters.
}
\label{fig:training-quintic-2}
\end{figure}

CYJAX presently allows for the study of single defining equations with multiple moduli, in which case the input to the neural networks includes multiple moduli. 
To illustrate this, we perform training for two moduli parameters with the training routine essentially as above\footnote{Using the same number and sizes of hidden layers. The powers used here are integers from $1$ to $6$. Note that the architectures here do not yet constitute the optimum over an extensive hyperparameter search.}. 
We consider the quintic defined by
\begin{equation}
Q_{\alpha_1,\alpha_2}(z)=\sum_{i=0}^{4} z_i^{5} +\alpha_1 \prod_{i}z_i + \alpha_2 \sum_{i \neq j}z_i^4 z_j = 0 \,.
\label{eq:2moduli}
\end{equation}
The moduli parameters are again uniformly sampled with real and imaginary parts ranging from $-10$ to $10$.
The training behavior and the final accuracy are summarised in Figure~\ref{fig:training-quintic-2} for degree $k=4$. This example demonstrates that training is still converging, but the accuracy seems to be worse than in the one modulus case. 
This is not surprising as the problem becomes more difficult the more moduli parameters we consider.
A systematic analysis and a more elaborate hyper-parameter scan for improved networks is beyond the scope of these notes.

\subsection{Comparison with Donaldson's algorithm for the Dwork quintic}
\begin{figure}
\begin{center}
\includegraphics[width=0.55\textwidth]{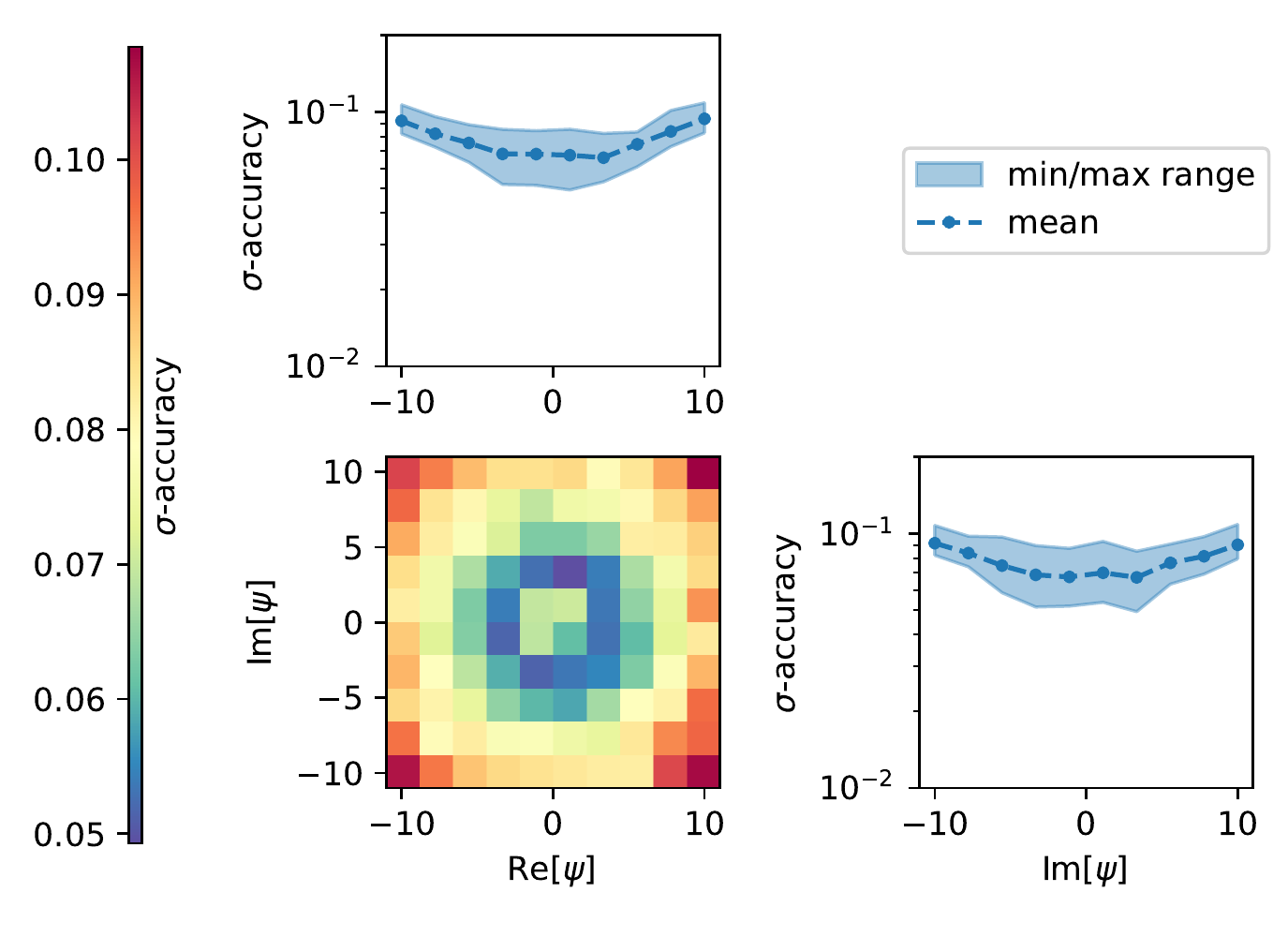}
\end{center}
\vspace{-10pt}
\caption{Accuracy for algebraic metrics obtained by running Donaldson's algorithm individually at different values of moduli. The results are shown for $k=6$ and using $15$ iterations, each comprising an integral approximated with $95000$ sample points.}\label{fig:donaldson}
\end{figure}

When evaluating neural network metrics, it is a good sanity check to compare with other methods for obtaining metrics. 
One such method is to construct metrics using Donaldson's algorithm. 
The CYJAX package implements methods to compute these and a notebook showing how to run it can be found in the documentation. 
Here we show an example of results obtained for evaluating it for the Dwork quintic at different points in moduli space at $k=6$. 
The results are shown in Figure~\ref{fig:donaldson}. 
We observe that the qualitative behavior of the accuracies resemble the ones obtained with our neural networks as shown in Figure~\ref{fig:training-dwork}.

Averaged over the selected range of moduli values, the $\sigma$-accuracy measure achieved with machine learning is about $4.4$ times smaller (i.e.~better) than the one with Donaldson's algorithm.
Note, however, that it is well known that Donaldson's algorithm does not yield the optimal metric at each fixed degree $k$ \cite{Headrick:2009jz}.

\section{Outlook}
\label{sec:outlook}
As described in the beginning, these notes are intended to describe only the initial release of CYJAX. There are many ways we intend it to be utilized. 
Below are some possible directions to extend it:
\begin{itemize}
\item One natural step for extension is to include other varieties, where complete intersection Calabi-Yau manifolds~\cite{Candelas:1987kf,Green:1987cr} and the Kreuzer-Skarke list~\cite{Kreuzer:2000xy} are natural candidates. As more manifolds become available, it will be natural to connect with other packages focusing on different aspects such as CYTOOLS~\cite{Demirtas:2022hqf}.
\item Another natural direction is to incorporate more general networks such as learning the metric directly or using different ans\"atze for the K\"ahler potential.
\item Many components of CYJAX can be reused for the implementation of more general metrics, such as metrics with $SU(3)$ structure as demonstrated in~\cite{Anderson:2020hux}.
\item We hope that the inherently auto-differentiable structure of the metric enables the investigation of other geometric objects which appear for phenomenologically relevant properties. For instance our metric is differentiable with respect to the moduli which is inherently necessary to link with studies of moduli stabilization.
\end{itemize}

Apart from these methodological extensions, there is a natural quest for achieving networks which result in high-accuracy metrics. As illustrated in our two moduli example (cf.~Section~\ref{sec:twomoduli}), there is room to improve upon our current illustrative networks to obtain better approximations to the metric. We hope to return to many of these aspects in the not too distant future.

\section*{Acknowledgments}
We thank Piotr Kucharski and Davide Passaro for extensive discussions and collaboration on related topics.
We acknowledge the Mainz Institute for Theoretical Physics (MITP) of the Cluster of Excellence PRISMA+ (Project ID 39083149) for hospitality and support during part of this work.

\bibliographystyle{inspire}
\bibliography{references}
\end{document}